# Toughening elastomers via microstructured thermoplastic fibers with sacrificial bonds and hidden lengths


Shibo Zou (邹士博) [a*], Daniel Therriault [a], Frédérick P. Gosselin [a]

[a] Laboratory for Multiscale Mechanics (LM²), Department of Mechanical Engineering, Research Center for High Performance Polymer and Composite Systems (CREPEC), Polytechnique Montréal, Montréal, QC H3T 1J4, Canada.

[*] Corresponding author. Email: shibo.zou@polymtl.ca



**Abstract:** Soft materials capable of large inelastic deformation play an essential role in high-performance nacre-inspired architectured materials with a combination of stiffness, strength and toughness. The rigid "building blocks" made from glass or ceramic in these architectured materials lack inelastic deformation capabilities and thus rely on the soft interface material that bonds together these building blocks to achieve large deformation and high toughness. Here, we demonstrate the concept of achieving large inelastic deformation and high energy dissipation in soft materials by embedding microstructured thermoplastic fibers with sacrificial bonds and hidden lengths in a widely used elastomer. The microstructured fibers are fabricated by harnessing the fluid-mechanical instability of a molten polycarbonate (PC) thread on a commercial 3D printer. Polydimethylsiloxane (PDMS) resin is infiltrated around the fibers, creating a soft composite after curing. The failure mechanism and damage tolerance of the composite are analyzed through fracture tests. The high energy dissipation is found to be related to the multiple fracture events of both the sacrificial bonds and elastomer matrix. Combining the microstructured fibers and straight fibers in the elastomer composite results in a ~ 17 times increase in stiffness and a ~ 7 times increase in total energy to failure compared to the neat elastomer. Our findings in applying the






sacrificial bonds and hidden lengths toughening mechanism in soft materials at the microscopic scale will facilitate the development of novel bioinspired laminated composite materials with high mechanical performance.

## 1. Introduction

Nature often assembles materials with intricate architectures, achieving functional properties that are superior than the intrinsic properties of their constituents [1–4]. For example, nacre from mollusk shells, a composite of 95 vol% minerals and 5 vol% organic materials, is 3000 times more fracture resistant than the mineral component.[5] The "brick-and-mortar" architecture with microscale tablets of calcium carbonate bonded by nanoscale interlayer of chitin and protein in nacre leads to an attractive combination of stiffness, strength and toughness, which has inspired the development of advanced composite materials [6–10]. Soft organic materials, despite their small amounts in nacre, play a critical role in spreading nonlinear deformations over large volumes and achieving high toughness at the macroscopic scale [11]. Saw-tooth patterns were observed on the force-extension curves of the soft organic materials in nacre [12], attributing to the breaking of sacrificial bonds and unfolding of hidden lengths at the molecular scale [13]. The sacrificial bonds and hidden lengths toughening mechanism is also found in bone [14], spider silk [15] and mussel byssus threads [16]. Recent studies [6–10] on bioinspired composites often focus on duplicating the architecture in nacre at the microscopic scale to achieve crack deflection and bridging or interfacial sliding toughening mechanisms, while few [17,18] focuses on the development of soft interface materials that mimic the large inelastic deformation and high energy dissipation of natural organic materials. The challenge lies in reproducing the sacrificial bonds and hidden lengths toughening mechanism in engineering materials.





Synthetic elastomers have a similar large deformation behavior to the soft organic materials in nacre. Previous studies [6–10] using elastomers as the soft interface in nacre-inspired architectured composites successfully improved the crack growth resistance, damage tolerance and impact resistance. Cavities and ligaments are formed at the elastomer interface under large deformation [6], which is consistent with the behavior of nacre. However, the deformation of elastomers is mostly elastic and thus recoverable, while the soft organic interface in natural materials shows large inelastic deformations and high energy dissipation. Strategies like introducing sacrificial bonds and folded domains into elastomers at the molecular scale [19,20] are effective to achieve large inelastic deformations and improve the energy dissipation in synthetic elastomers. Researchers also explored the possibility of reinforcing elastomers with rigid mesh at the microscopic or macroscopic scale to achieve the large inelastic deformation and high energy dissipation [21–23]. A saw-tooth tensile behavior was accomplished through the multiple fracture of the rigid mesh inside the elastomer matrix. Feng et al. [21] incorporated a polyamide fabric mesh in between acrylic tapes with viscoelastic properties. The interlayer sliding induces shear stresses between the mesh and tape, which fracture the mesh into multiple islands until a critical length scale is reached. Their resulting composite shows a ~ 5 times increase in stiffness and a ~ 2 times increase in total energy to failure compared to the acrylic tape. King et al. [23] embedded a 3D-printed plastic mesh inside an elastomer matrix. The multiple fracture of the plastic mesh triggered by a topological interlocking mechanism leads to a ~ 60 times increase in stiffness and ~ 0.5 time increase in total energy to failure compared to the neat elastomer. Both examples adopted the double network strategy [24,25] which uses rigid mesh as the first network to ensure high stiffness and energy dissipation, and soft matrix as the second network to maintain the structural integrity. Since the soft matrix is still intact after the rigid mesh fractures, the





deformation of the composite is recoverable. The energy dissipation can be further increased by introducing molecular-level sacrificial bonds into the matrix [22]. Recently, we demonstrated large inelastic deformation and high energy dissipation in elastomer composites under low-velocity impact [26] by combining the widely used elastomer PDMS and the microstructured thermoplastic fibers with sacrificial bonds and hidden lengths [27,28].

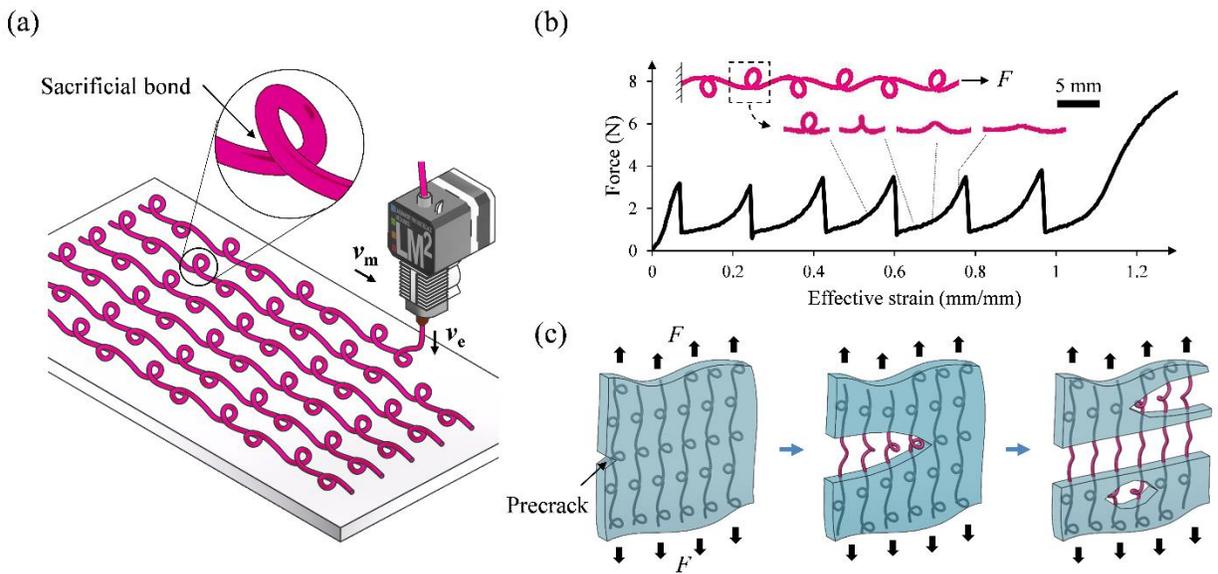

Figure 1. Overview of toughening elastomers via microstructured thermoplastic fibers with sacrificial bonds and hidden lengths. (a) Schematic illustration of the IFFF process. The extruding speed $v_e$ is 2.06 times faster than the extruder moving speed $v_m$, producing the alternating pattern. (b) Force-effective strain curve of the as-fabricated microstructured fibers with six alternating loops. The breaking of sacrificial bonds and the unfolding of hidden lengths from one of the six loops are illustrated as an example. The fiber contours with fake color are traced from camera snapshots during the tensile test of the microstructured fiber. (c) Schematic illustration of multiple breaking of sacrificial bonds and multiple cracking of the matrix in an elastomer composite with a precrack under tension.





The microstructured fibers with sacrificial bonds and hidden lengths are made by instability-assisted fused filament fabrication (IFFF) as shown in Fig. 1a. A molten polymer thread falls from a certain height onto a platform with a relative movement in the horizontal direction. Depending on the ratio of the extruding speed $v_e$ to the printing head moving speed $v_m$, a series of nonlinear patterns can be produced due to the fluid-mechanical instability [29]. In the alternating pattern that we chose for this study, the self-intersections of the molten polymer thread form weakly fused bonds after cooling, which act as sacrificial bonds under mechanical loading. Fig. 1b shows a representative force-effective strain curve of an alternating PC fiber (dia. = 0.44 mm) with six loops under tensile loading. Each force peak corresponds to the breaking of a sacrificial bond. After the bond breaking, the hidden length, i.e., the fiber loop, is released and unfolded. The unfolding of hidden lengths leads to large-scale plastic deformation along the fiber [30]. Embedding the microstructured fibers in PDMS produces a transparent elastomer composite with graceful failure and superior damage tolerance [26]. Although large inelastic deformation and high energy dissipation were demonstrated in the elastomer composite under impact loading [26], the composite's fracture mechanism and damage tolerance under static loading remain to be investigated. In this work, we perform uniaxial tensile tests of the alternating fiber composite specimens in the pure shear geometry [31] with and without a precrack to understand the interactions between crack propagation, bond breaking and fiber unfolding. Two levels of multiple fracture events are observed during tensile tests: the multiple breaking of sacrificial bonds along the fiber and the multiple cracking of the PDMS matrix. The breaking of sacrificial bonds, unfolding of fiber loops and the formation of voids in PDMS dissipate the mechanical energy which would otherwise drive the propagation of a precrack (Fig. 1c). The voids and exposed fibers are analogous to the micro cavities and ligaments at the soft organic interface of nacre. We also





compare the total energy to failure and damage tolerance between the plain elastomer and its composites with different fiber reinforcements under uniaxial tensile loading. In the end, static puncture tests are carried out to evaluate the energy absorption and damage tolerance of the composite under transverse loading.

## 2. Materials and methods

2.1 Microstructured Fiber Printing

We adopted the instability-assisted fused filament fabrication technique [28] to make microstructured fibers with sacrificial bonds and alternating loops. A PC filament (Top3d Filament, Dongguan, China) with a diameter $D_{filament} = 1.75$ mm was dried overnight in a vacuum oven at 65 °C before printing on a Prusa i3 printer with a nozzle diameter $d_{nozzle} = 0.4$ mm at 330 °C. We used the Simplify3D software to send the custom-written g-code to the printer. In the g-code, we chose a deposition height of 5 mm for the whole printing process to keep the deposition within the viscous steady coiling regime [32]. In order to produce the alternating pattern, the speed ratio $v_e/v_m$ should be in the range of 1.57 to 2.23 according to the geometric model developed by Brun et al. [33]. The horizontal moving speed $v_m$ is directly determined by the printing head moving feedrate $V_F$ from the g-code. Based on the continuity equation for incompressible fluids, the extruding speed $v_e$ is determined by the following equation:

$$C_1 \frac{\pi}{4} D_{filament}^2 \cdot \frac{L_E}{(L_D/V_F)} = \frac{\pi}{4}(C_2 d_{nozzle})^2 \cdot v_e, \tag{1}$$

$$\text{i.e., } v_e = \frac{C_1 L_E V_F}{L_D}\left(\frac{D_{filament}}{C_2 d_{nozzle}}\right)^2, \tag{2}$$

where, the extruding length $L_E$ of the 1.75 mm filament, the moving distance $L_D$ of the printing head, and the printing head moving feedrate $V_F$ are independent parameters in the g-code. We first





fixed $V_F$ at 1929.3 mm/min. Then we measured the volume flow rate correction factor $C_1$ (0.82)

and the fiber expansion factor $C_2$ (1.09) experimentally [26]. The speed ratio $v_e/v_m$ is proportional

to the ratio of two independent g-code parameters $L_E/L_D$:

$$r = v_e/v_m = \frac{C_1 L_E}{L_D}\left(\frac{D_{\text{filament}}}{C_2 d_{\text{nozzle}}}\right)^2. \tag{3}$$

We chose a speed ratio $v_e/v_m = 2.06$ to produce the alternating pattern in this work. Therefore, for

a target printing distance $L_D$, the extruding length $L_E$ can be calculated by equation (3). The

horizontal printing direction was kept the same for all fibers. The fiber separation was defined in

the g-code. After the printing of all fibers, a rectangular frame was printed on top of the fibers for

the easy handling of the fibers without altering the distance between each fiber. The straight fibers

were printed using the same g-code for alternating fibers in which the deposition height was

changed to 0.4 mm. Therefore, the straight fiber has the same weight as the alternating fiber.

## 2.2 PDMS-Microstructured Fiber Composite Fabrication

Polydimethylsiloxane (Sylgard 184, Dow) was used to make all test specimens. The prepolymer

base and the curing agent were thoroughly mixed at a ratio of 10:1, and then degassed in a vacuum

oven at room temperature for one hour. Two acrylic plates were laser-cut and assembled into a

rectangular mold with a chamber size of $100 \times 80 \times 1.5$ mm (length $\times$ width $\times$ depth). Before

placing the as-printed fibers into the mold, two sides of the rectangular frame which are parallel to

the fibers were cut, and the other two sides were cast in PDMS together with the fibers. Hence, the

fibers stay aligned during the molding process. Any bubbles generated during casting would

disappear after a short period in ambient environment. Then we put the mold into an oven and

cured the mixture at 65 °C for 4 h. The as-fabricated specimen had a thickness of 1.5 mm. We used

the same process to make the plain PDMS specimen and PDMS-unidirectional fiber composite





specimens. We first prepared the PDMS-alternating fiber composite specimen with 23 evenly distributed fibers and a fiber separation of 4.35 mm for the failure mechanism and crack growth resistance analysis of the composite. In order to compare the tensile behavior of different composite specimens, we then prepared the PDMS-alternating fiber and PDMS-straight fiber composite specimens with 12 evenly distributed fibers and a fiber separation of 8.7 mm, and the PDMS-hybrid fiber composite specimen with 12 alternating fibers, 11 straight fibers and a fiber separation of 4.35 mm. For the PDMS-hybrid fiber composite specimen, the alternating fibers were manually aligned to be in the middle of the straight fibers in the mold before pouring the PDMS resin. The PDMS-bidirectional fiber composite specimens for the static puncture test all have a size of $100 \times 100 \times 1.5$ mm (length $\times$ width $\times$ thickness) and were fabricated using the procedure described in reference [26].

## 2.3 Mechanical Testing

We conducted all the mechanical tests on an MTS Insight electromechanical machine with a 100N or 1000N load cell depending on the test load range. Tensile test of the alternating fiber with six loops was carried out with a gauge length of 32 mm and crosshead speed of 500 mm/min. The pure shear specimens have a length of 100 mm and a thickness of 1.5 mm. The grips opening is 20 mm. The crosshead speed is 10 mm/min for all pure shear specimens. We glued the pure shear specimens to laser-cut PC grips with 3M silicone sealant 8661. Mounted screws were also used to tighten the grips in order to make sure that there is no slippage of the specimen. We used scissors to create the precrack with a length of 30 mm on the test specimen. Any fibers along the precrack were also cut. The crack extension of the notched plain PDMS specimen was measured by a custom-written MATLAB code which can detect the crack tip movement based on digital image analysis. The crack extension of the notched PDMS-alternating fiber composite specimen was





manually measured in Image J. For a purely elastic material, the energy release rate $G$ can be calculated by the following equation [31]:

$$G = W(\lambda)H, \tag{4}$$

where $H$ is the height of the undeformed specimen, $\lambda = 1 + \varepsilon = 1 + d/H$ is the stretch ratio, $d$ is the crosshead displacement, $\varepsilon$ is the effective strain, $W(\lambda)$ is the strain energy density of the material ahead of the crack tip measured by integrating the area under the stress-strain curve of an unnotched sample with identical dimensions [34,35]. The energy release rate of hyperelastic materials like PDMS can be directly calculated by equation (4) [36]. A critical energy release rate $G_c$ was calculated at the onset of crack growth with a critical stretch ratio $\lambda_c$. The $G_c$ was taken as the fracture toughness according to previous studies [25,37,38]. Since equation (4) has also been used to analyze materials with large scale inelasticity [35], we used equation (4) to calculate the effective energy release rate and effective fracture toughness of our dissipative PDMS-alternating fiber composite, in order to compare with the neat elastomer. The static puncture test was carried out with a hemispherical indenter head (diameter: 25.4 mm). The specimen was clamped by two acrylic plates (thickness: 12.7 mm) with a circular opening (diameter: 76.2 mm). Mounted screws were used to tighten the plates and avoid the slippage of the specimen inside the clamps. The indenter speed was 10 mm/min for all static puncture test specimens. The total energy to failure of both pure shear test specimens and static puncture test specimens were calculated by integrating the area under the force-displacement curves.

## 3. Results and Discussion





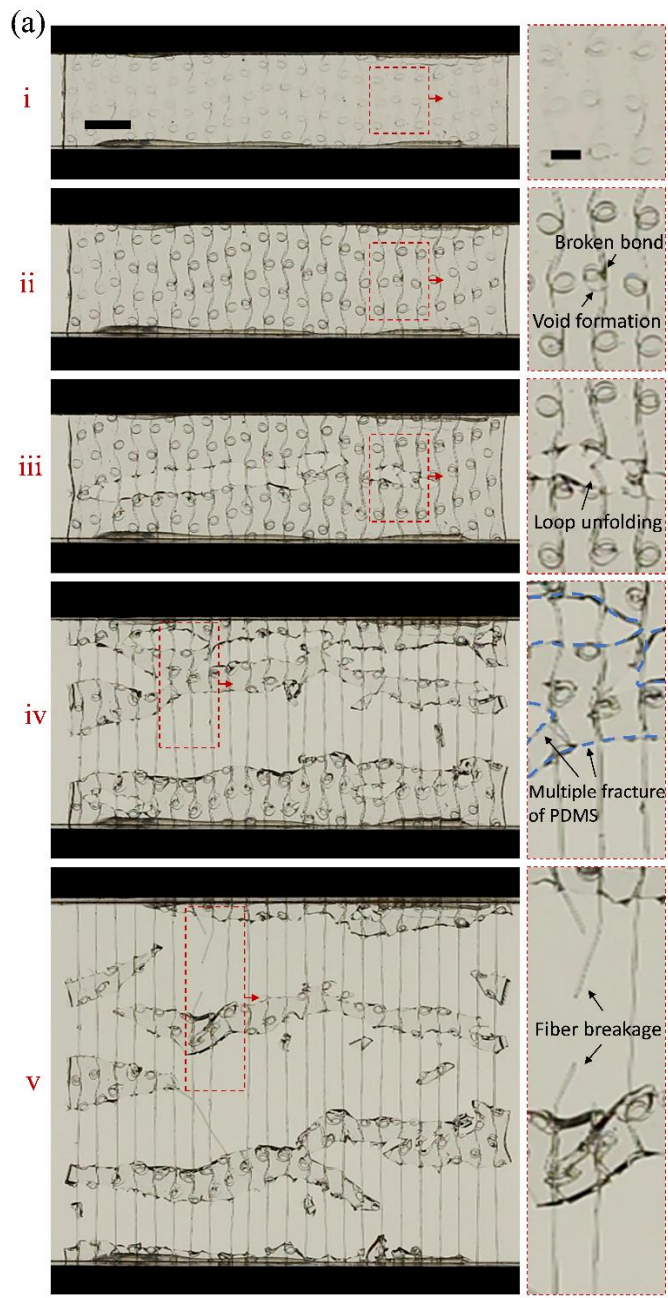

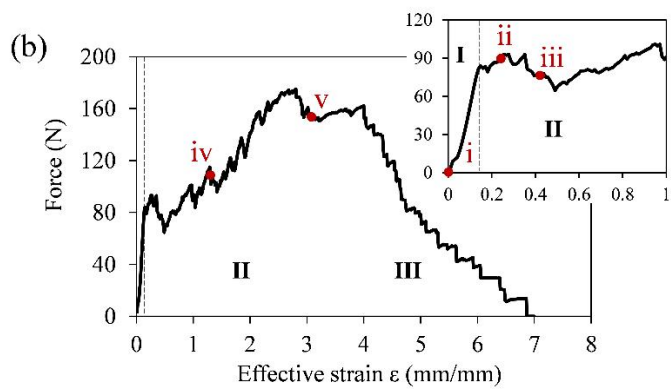





Figure 2. Fracture of the unnotched PDMS-alternating fiber composite specimen under uniaxial tension. (a) Camera snapshots of the fracture of the specimen at different effective strains: i. $\varepsilon = 0$; ii. $\varepsilon = 0.24$; iii. $\varepsilon = 0.41$; iv. $\varepsilon = 1.32$; v. $\varepsilon = 3.04$. Scale bar is 10 mm in the camera snapshot and 3 mm in the inset. (b) Force-effective strain curve of the PDMS-alternating fiber composite.

We first tested a PDMS-alternating fiber composite specimen without a precrack (Fig. 2). Three regimes were observed from the test: I. homogeneous stretching of the fiber and PDMS; II. breaking of sacrificial bonds inside PDMS, void formation in PDMS, unfolding of the released loop and segmentation of PDMS; III. Onset of fiber breakage and final failure of the specimen. Due to the similar refractive indexes of the PC fiber and PDMS [26], the fiber is almost invisible inside PDMS when the specimen is at rest (see Fig. 2a i). Once the specimen is stretched, the nonuniform deformations induce the detaching of the fiber and matrix, making the fiber contour more apparent. Fig. 2b shows that the force linearly increases with respect to the effective strain in regime I. In regime II, the force-effective strain curve first reaches a plateau (see Fig. 2b ii, $\varepsilon = 0.24$ mm/mm) due to the breaking of sacrificial bonds inside the matrix. Since the adhesion between the PC fiber and PDMS is weak, the loop unfolds inside the matrix as the stretching continues, resulting in fiber-matrix sliding. The unfolding of the loop enlarges the void inside the matrix (see Fig. 2a ii). The void acts like an embedded crack and propagates transversely through the thickness of the specimen. Once the embedded crack becomes a through crack, the fiber is fully exposed, and the loop is free to unfold (see Fig. 2a iii). Then the crack propagates longitudinally in PDMS and exposes more loops. The multiple breaking of sacrificial bonds inside the matrix initiates more voids that evolve into running cracks. The interlocking between the fiber loop and the matrix triggers the multiple fracture of PDMS (see Fig. 2a iv). Similar fracture behavior was observed by Chang et al. [39] in PDMS-polyamide fabric composite. The detaching





of fiber and PDMS makes the voids into running cracks under large deformation. In Chang et al.'s work, after matrix cracking, the polyamide fabric goes through a bending to stretching transition as the stretching continues [39]. In our case, the loop first unfolds after matrix cracking with a coupled deformation of bending, torsion and axial tension [40]. As the stretching continues, the straightened fiber yields under tension. The strain hardening of the fibers results in the force increase in regime II. In regime III, the fibers break one by one (see Fig. 2a v), leading to the gradual failure of the specimen.

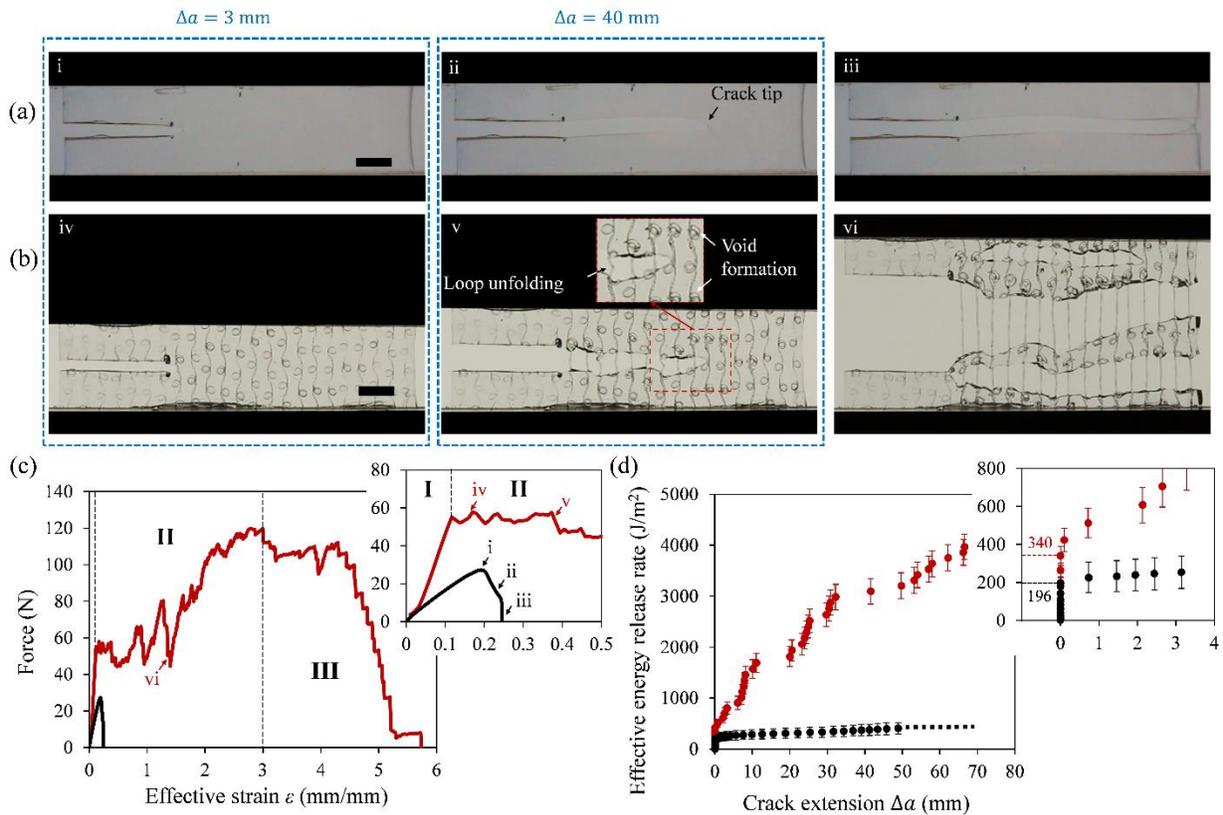

Figure 3. Comparison of the fracture behavior of the notched plain PDMS specimen and the notched PDMS-alternating fiber composite specimen under tensile loading. (a) Camera snapshots of the notched plain PDMS specimen at different effective strains: i. $\varepsilon = 0.19$; ii. $\varepsilon = 0.23$; iii. $\varepsilon = 0.25$. (b) Camera snapshots of the notched PDMS-alternating fiber composite specimen at different





effective strains: iv. $\varepsilon = 0.17$; v. $\varepsilon = 0.38$; vi. $\varepsilon = 1.38$. All scale bars are 10 mm. (c) Force-effective strain curves of the notched plain PDMS (in black) and PDMS-alternating fiber composite (in red) specimens. (d) Crack growth resistance curves of the notched plain PDMS (in black) and PDMS-alternating fiber composite (in red) specimens. The error bar represents the standard deviation of the effective energy release rate with the strain energy density measured from three unnotched specimens with identical dimensions as the notched specimen. In the notched plain PDMS specimen, when the crack extension exceeds 50 mm, the crack propagation accelerates rapidly, making it difficult to measure the crack extension from the camera snapshots. Since the effective energy release rate stays almost at the same level, we extrapolated the dashed line after the crack extension exceeds 50 mm. The inset shows the critical effective energy release rate of the plain PDMS ($196 \pm 74$ J/m$^2$) and PDMS-alternating fiber composite ($340 \pm 49$ J/m$^2$) at the onset of crack growth.

The low fracture resistance of PDMS has been previously reported [38,39]. Fig. 3a shows the fracture of a notched PDMS specimen in the pure shear geometry. The crack starts to propagate at $\varepsilon = 0.18$ and runs through the whole specimen very quickly. The mechanical behavior of a notched PDMS-alternating fiber composite specimen also includes three regimes (Fig. 3b and c). The fiber and PDMS are co-stretched in regime I. The breaking of sacrificial bonds and the fiber-matrix sliding produce voids in regime II. With the voids developing into running cracks, the fiber is exposed and straightened, and the specimen is segmented into several isolated PDMS islands. Finally, the specimen fails after the breaking of fibers in regime III. The precrack of the PDMS-alternating fiber composite specimen starts to propagate at $\varepsilon = 0.12$ in regime II. In the notched PDMS specimen, the energy dissipation happens only at the crack tip, via breaking of molecule





chains. While in the notched PDMS-alternating fiber composite specimen, more energy dissipation mechanisms are introduced both behind and ahead of the crack tip and are discussed below.

In Fig. 3d, the energy release rate $G_{PDMS}$ calculated for the notched plain PDMS specimen represents the driving force for crack growth per unit area $dA$. The $G_{PDMS}$ at the onset of crack propagation ($196 \pm 74$ J/m$^2$, Fig. 3d inset) is equal to the intrinsic fracture toughness $\Gamma_{PDMS}$ of the material, assuming that PDMS is purely elastic and incompressible. In the notched PDMS-alternating fiber composite specimen, the as-calculated effective energy release rate $G_{comp} = G_{tip} + G_{dis}$, where, $G_{tip}$ is associated with the actual amount of energy released to drive the crack to propagate per unit area $dA$, $G_{dis}$ is associated with the amount of energy that is first stored in the fiber and PDMS and then mostly dissipated by the breaking of sacrificial bonds, unfolding of hidden loops, growing of voids and fiber-matrix sliding during crack propagation per unit area $dA$. With the assumption that the fracture process is rate independent, $G_{tip}$ is equal to the intrinsic fracture toughness $\Gamma_{PDMS}$ and remains constant during crack propagation. Therefore, $G_{dis}$ during crack propagation can be calculated as $G_{comp} - \Gamma_{PDMS}$.

The effective energy release rate $G_{comp}$ at the onset of crack growth is $340 \pm 49$ J/m$^2$ in the PDMS-alternating fiber composite (Fig. 3d inset), which is higher than $\Gamma_{PDMS}$, but much lower than the critical energy release rate of tough elastomers ($1000 - 10000$ J/m$^2$) with molecular sacrificial bonds [38]. The relatively low $G_{comp}$ is due to the larger scale and smaller volume fraction of the sacrificial bonds and hidden lengths in our composite compared to the molecular sacrificial bonds in tough elastomers. The microstructured sacrificial bonds and hidden lengths in our composite contribute to high energy dissipation only at larger deformations. In the notched PDMS-alternating fiber composite specimen, the breaking of sacrificial bonds starts at $\varepsilon = \sim 0.12$, which is almost at the same time as the onset of crack propagation, meaning that the $G_{dis}$ at the onset of crack





propagation is mostly associated with the tensioning of the fiber, the fiber-matrix detaching and the breaking of a few sacrificial bonds. The resulting $G_{dis} = G_{comp} - \Gamma_{PDMS} = 144$ J/m$^2$, which is even less than $\Gamma_{PDMS}$. It should be noted that the tensioning of the fiber at early stretching stage contributes to $G_{dis}$ by storing elastic strain energy in the fiber, which is similar to the elastic dissipators in the literature [38,41], but these elastic strain energy will finally be dissipated by the breaking of sacrificial bonds and the unfolding of fiber loops during later stretching. At $\varepsilon = 0.17$, the precrack extends 3 mm (Fig. 3b iv), more sacrificial bonds are broken compared to the onset of crack propagation, but the fiber loops remain folded inside PDMS. The resulting $G_{dis} = G_{comp} - \Gamma_{PDMS} = \sim 500$ J/m$^2$. At $\varepsilon = 0.38$, the precrack extends 40 mm (Fig. 3b v), the fiber loops unfold behind the crack tip and voids form due to the fiber-matrix sliding ahead of the crack tip. The resulting $G_{dis} = G_{comp} - \Gamma_{PDMS} = \sim 2900$ J/m$^2$. Hence, the unfolding of fiber loops and voids formation happen at large effective strains and contribute to higher energy dissipation compared to the breaking of sacrificial bonds and fiber-matrix detaching at small effective strains. Even though several mechanisms are coupled during crack propagation, our observation is still consistent with previous finding [13] from entropic elasticity-based molecular simulations that the energy to unfold the hidden length is much higher than the energy to break a sacrificial bond. It should be noted that while dissipating energy, the formation of voids results in flaws with a length up to 2 mm which are larger than the fractocohesive length of PDMS at the order of magnitude of $10^{-1}$ mm [38,42]. On one hand, the voids increase the flaw-sensitivity of PDMS, which could be another reason for the low critical $G_{comp}$ in our composite. On the other hand, the development from voids into running cracks facilitates the exposing and unfolding of the fiber loops, leading to extra energy dissipation after the matrix cracking.





In tough elastomers with molecular sacrificial bonds, both experimental [19] and analytical [35,43] results show that sacrificial bonds break at the crack tip, leading to an energy dissipation zone with a strip shape surrounding the crack surface. The loading and unloading of the material in this zone increase the energy dissipation during crack propagation. In our notched PDMS-alternating fiber composite, the sacrificial bonds break and voids form in a wide region between the clamps ahead of the crack tip, which is equivalent to intrinsic toughening [44]. Once the crack propagates, the fiber behind the crack tip is unloaded due to the exposing of the fiber loop. As the fiber loop continues to unfold, the large-scale plastic deformation [30] along the fiber increases the energy dissipation behind the crack tip, which is equivalent to extrinsic toughening [44]. Even though the energy dissipation zone seems larger in our composite compared to the strip shape zone in tough elastomers with molecular sacrificial bonds, the volume fraction of sacrificial bonds in our composite is believed to be much lower than that in tough elastomers. The effective fracture toughness of the composite, i.e., the $G_{comp}$ at the onset of crack propagation, is not expected to significantly increase with the composite size, since the $G_{dis}$ at the onset of crack propagation is mostly associated with the tensioning of the fiber, the fiber-matrix detaching and the breaking of a few sacrificial bonds. However, the total energy to failure of the notched composite is expected to increase with the composite size. Because after the precrack cuts through the composite specimen longitudinally at $\varepsilon = 0.48$, the fiber continues to carry the load via plastic deformation. Further stretching develops more voids into running cracks (Fig. 3b vi), dissipating more energy till the final failure of the specimen. A larger composite size would lead to more energy dissipation before final failure. In order to take into consideration the amount of energy dissipation after the precrack cuts through the composite specimen in our analysis, we calculated the total energy to





failure instead of the effective energy release rate of different specimens in the following investigations.

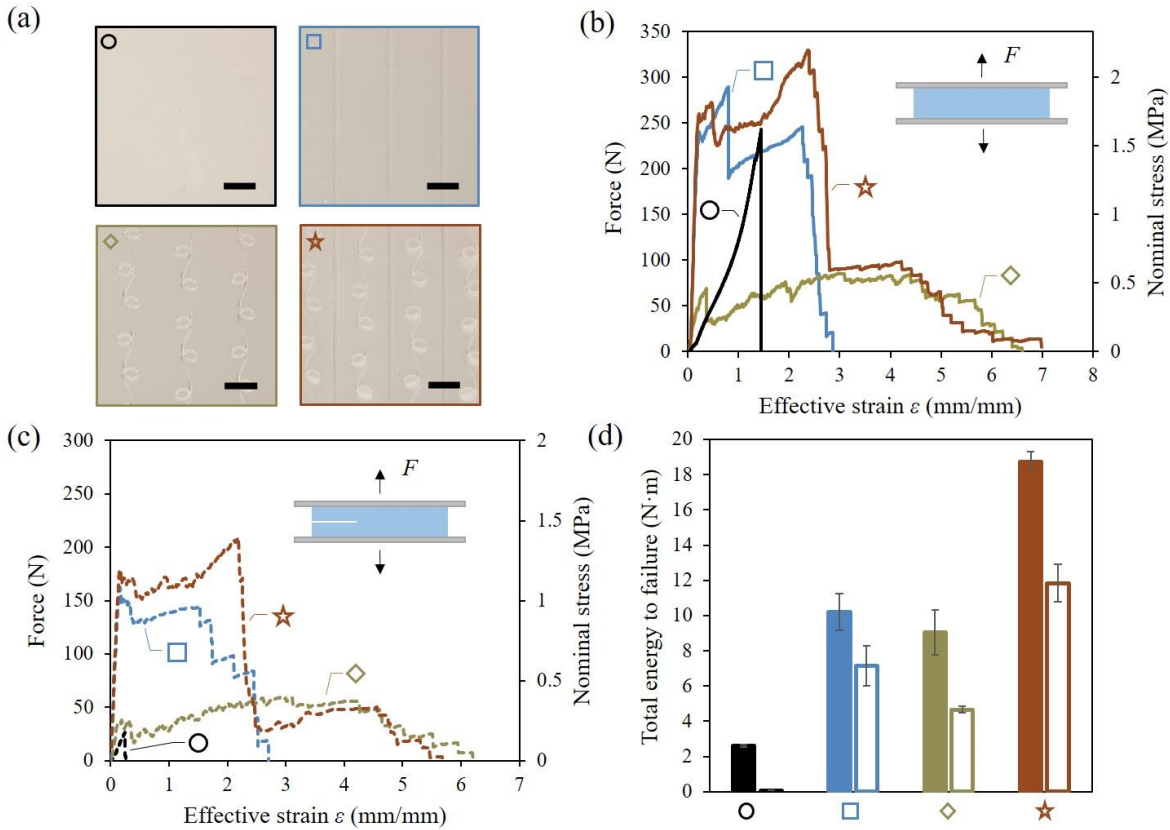

Figure 4. Tensile test of the plain PDMS specimen and PDMS-unidirectional fiber composite specimens with and without a precrack. (a) Optical images of test specimens: plain PDMS (○), PDMS-straight fiber composite (□), PDMS-alternating fiber composite (◇), PDMS hybrid fiber composite (★). (b) Force-effective strain curves of unnotched specimens. (c) Force-effective strain curves of notched specimens. The corresponding nominal stress is shown as a secondary vertical axis on the right. (d) Comparison of total energy to failure of test specimens. Solid blocks represent unnotched specimens, while hollow blocks represent notched specimens. All scale bars are 5 mm.





To compare the difference in mechanical performance between the microstructured fiber and the straight fiber in the composite, we prepared four specimens in the pure shear geometry as shown in Fig. 4a: plain PDMS, PDMS-straight fiber composite, PDMS-alternating fiber composite, and PDMS-hybrid fiber composite which contains both straight fiber and alternating fiber. Fig. 4b shows that the unnotched plain PDMS specimen exhibits a stiffness around 118 N/mm/mm and fractures at $\varepsilon = 1.45$. The unnotched alternating fiber composite specimen shows the aforementioned three regimes with a stiffness around 272 N/mm/mm and final failure at $\varepsilon = 6.61$. The unnotched straight fiber composite specimen shows a linear regime until $\varepsilon = 0.20$ on the force-effective strain curve and a high stiffness around 1718 N/mm/mm. At $\varepsilon = 0.80$, a crack initiates at the straight fiber-matrix interface and ruptures through the whole matrix, leading to the force drop on the force-effective strain curve. Then the straight fibers continue to carry the load. The strain hardening of the straight fibers leads to an increase on the force-effective strain curve. At $\varepsilon = 2.27$, the fibers start to break one by one, resulting in the final failure of the specimen at $\varepsilon = 2.86$. Compared to the as-printed straight fiber, the average fiber width of the broken straight fibers reduced by 28.3%, indicating the fibers' plastic deformation. Whereas the strain at failure of the unnotched alternating fiber composite specimen is ~131% higher than that of the unnotched straight fiber composite specimen, the total energy to failure of the former is ~11% lower than that of the latter. We believe that when the material has an exceptional strain hardening behavior like polycarbonate, the advantage of toughness enhancement by the microstructured fibers with sacrificial bonds and hidden lengths compared to the straight fiber would be diminished, since the bond-breaking defects and bending-torsion-tension coupled loop deformation hinder the material from reaching the same amount of ultimate strain and strength as the straight fiber, resulting in the relatively lower energy absorption in microstructured fibers.





Combining both the alternating fibers and straight fibers in the hybrid fiber composite results in a force-effective strain curve which is almost the sum of that of the alternating fiber and the straight fiber composites, with a strain at failure ($\varepsilon = 6.98$) similar to the alternating fiber composite and a stiffness (2052 N/mm/mm) similar to the straight fiber composite. The notched plain PDMS specimen shows a large decrease in strength and strain at failure, while the notched composite specimens keep similar stiffness and strain at failure to their unnotched counterparts, with only a decrease in strength due to the cut of fibers along the precrack. The total energy to failure of the notched plain PDMS specimen is only 2.2% of that of the unnotched specimen, while the notched composite specimens keeps 50% – 70% of the total energy to failure of their unnotched counterparts. The high damage tolerance of the composites is mainly attributed to the plastic deformation of the alternating and/or straight fibers after matrix cracking.

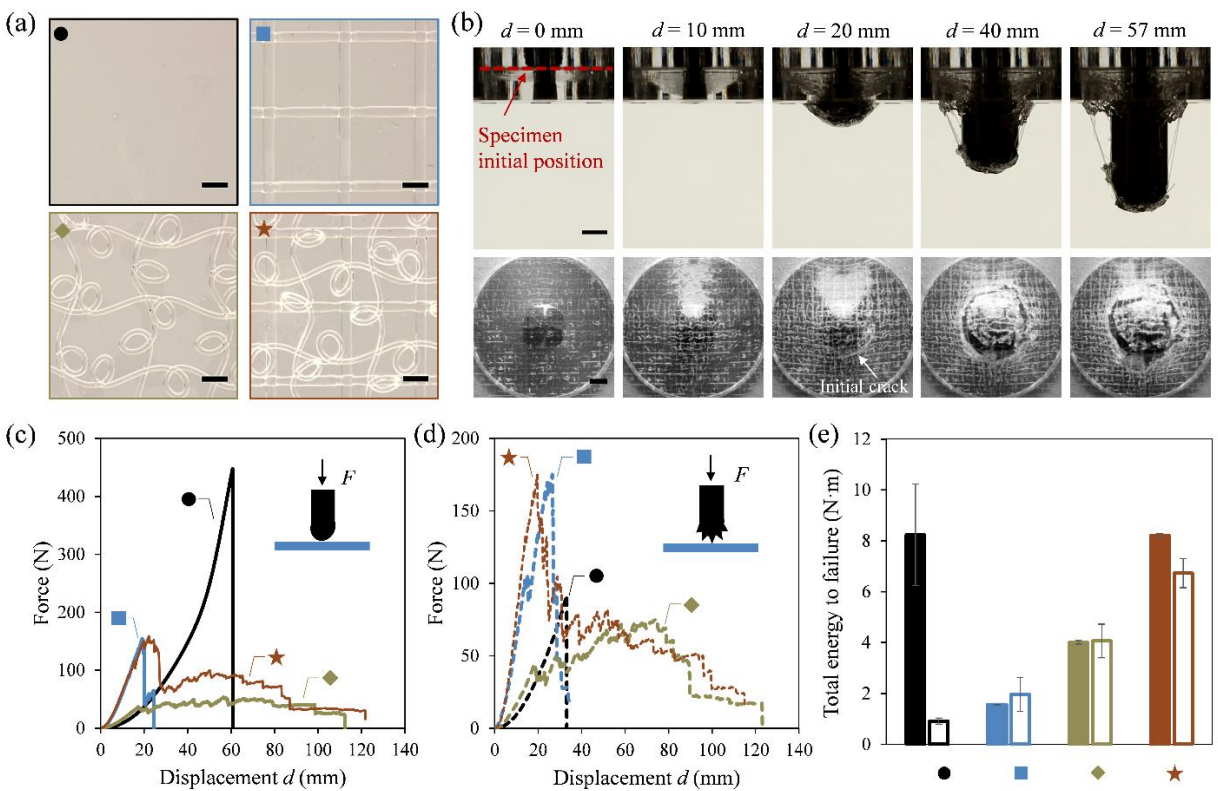





Figure 5. Static puncture test of the plain PDMS specimen and PDMS-bidirectional fiber composite specimens. (a) Optical images of test specimens: plain PDMS (●), PDMS-straight fiber composite (■), PDMS-alternating fiber composite (◆), PDMS-hybrid fiber composite (★). Scale bars are 2 mm. (b) Camera snapshots of the static puncture test of PDMS-hybrid fiber composite with a rough indenter head in front and bottom views. Scale bars are 10 mm. Force-displacement curves of test specimens with (c) a smooth indenter head, and (d) a rough indenter head attached with a P60 grade sandpaper. (e) Total energy to failure of test specimens. Solid blocks represent tests with the smooth indenter head, while hollow blocks represent tests with the rough indenter head.

We further evaluated the composite's energy absorption performance under transverse loading. Fig. 5a shows the four types of specimens: plain PDMS, PDMS-straight fiber composite, PDMS-alternating fiber composite, and PDMS-hybrid fiber composite. We first carried out the static puncture test with a smooth indenter head. The plain PDMS specimen endures a high transverse force up to 448 N. The strength of the plain PDMS here is higher than all other composite specimens, which is different from the comparison in the tensile test results in Fig. 4b, because the composites carry the transverse load with only a few fibers that are under the loading path of the puncture head, while all the fibers are loaded during the tensile tests in Fig. 4b. Under the transverse load, the plain PDMS specimen fails when a crack initiates at the edge of the indenter head and propagates rapidly and circularly around the indenter head. The failure of the plain PDMS specimen produces a popping sound with the remaining pieces bouncing up and down rapidly, indicating that a large amount of the energy absorption releases and transforms into sound and kinetic energy after unloading. The material failure is more graceful in the alternating and hybrid





fiber composite specimen (Fig 5b), with mechanical energies dissipated away instead of stored inside the material.

The alternating fiber composite specimen fails with a circular crack propagation similar to that of the plain PDMS specimen, but in a much slower pace. After the circular crack propagates fully around the indenter head, the alternating fibers continue to unfold and the interlocking between the alternating fiber and matrix leads to the multiple cracking of PDMS. In some cases, the alternating fiber cuts out from the matrix transversely, which are not observed in the tensile tests. The fiber cut-outs mostly happen at the bottom layer of the bidirectional fabric, where the alternating fibers are easier to cut through the matrix than the top layer fibers. Due to the limitation in clamping force with screws and the short clamping length (~ 12 mm) of the fiber in this study, after the cut-outs, the fibers often slide out of the matrix instead of breaking. Since the sliding of the fibers always happens before the final breaking of those fibers without cutting out from the matrix, we believe the sliding only affects the load-bearing capability, not the extensibility. In our tests, the breaking of those fibers without cutting out from the matrix results in the final failure of the alternating fiber composite.

After matrix cracking, the straight fiber composite specimen does not show large yielding of the fibers similarly to the results shown in Fig. 4. Conversely, the matrix cracks after the breaking of the straight fibers at the grid intersection point. The printing of the straight fiber fabric weakens the fibers at the intersections, hindering the strain hardening potential of the fibers. Fig. 5c shows that the loading capacities of the three composite specimens are 13% − 33% of that of the plain PDMS specimen, under current fiber volume fraction. The hybrid fiber composite specimen achieves almost the same total energy to failure as the plain PDMS specimen (Fig. 5e). Considering that only three alternating fibers and three straight fibers in each direction contribute to the





penetration resistance under the current indenter size, the loading capacity and energy absorption can be further improved with a higher fiber volume fraction.

To evaluate the specimen's damage tolerance under transverse loading, we attached a P60 grade sandpaper to the hemispherical indenter which introduces surface cracks to the specimen during loading. The maximum depth of the surface crack is $0.3 - 0.4$ mm (Fig. S1), which is comparable to the fractocohesive length of PDMS at the order of magnitude of $10^{-1}$ mm [38,42]. Compared to the test results with a smooth indenter head, the loading capacity and energy absorption of the plain PDMS specimen with a rough indenter head reduce by 80% and 89%, respectively (Fig. 5d-e). In contrast, the loading capacities and energy absorptions of the composite specimens with a rough indenter head are almost the same as these with a smooth indenter head. Compared to the alternating fiber composite, the alternating fiber fabric itself has very low load-bearing capability under puncture due to the weak adhesion between the two layers of the fabric which causes the slipping of the fiber around the indenter head. The high damage tolerance of the composites under transverse loading comes from the combination of the fiber fabric and the elastomer matrix, and various resulting energy dissipation mechanisms, including the plastic deformation of the fibers and the multiple fracture of the matrix, as well as the fiber-matrix sliding and fiber cut-outs.

## 4. Conclusions

We demonstrated large inelastic deformation and high energy dissipation under tensile and transverse loads in a commercially available elastomer reinforced by microstructured fibers with sacrificial bonds and hidden lengths. The deformation and failure of the composite material are dominated by the multiple breaking of sacrificial bonds at the microscale and the multiple fracture of the elastomer matrix at the macroscale. During these multiple fracture events, the composite material dissipates energy by the breaking of sacrificial bonds, yielding and fracture of the





unfolding fiber, voids formation and crack growth in the matrix, as well as fiber-matrix sliding. We show that the mechanical properties of the composite material can be tuned by controlling the pattern and arrangement of the reinforcing fibers. The mechanical performance of the composite may also depend on the fiber-matrix adhesion and matrix properties which will deserve further investigations. Instability-assisted 3D printing [27,28,45,46], electrospinning [47–49] and melt electrowriting [50,51] provide the opportunity for the miniaturization of the microstructured fiber to accommodate different indenter sizes in real-world applications. The transparency of the composite materials can be improved by matching the refractive indexes of the fiber and the elastomer matrix for applications that require high optical transparency [26]. The concept that we demonstrated here could be applied to any soft materials including elastomers and hydrogels in order to achieve large inelastic deformation and high energy dissipation. Combining our composite with other successful implementations of nature's toughening mechanisms in the literature [8–10] is a promising way to make advanced laminate composites with high energy dissipations at multiple length scales.

**Declaration of competing interest**

The authors declare that they have no known competing financial interests or personal relationships that could have appeared to influence the work reported in this paper.

**Acknowledgments**

We acknowledge the support of the Fonds de Recherche du Québec: Nature et Technologies (FRQNT), [funding reference number 63014], the Natural Sciences and Engineering Research Council of Canada (NSERC), [funding reference number 175791953], and the Canadian Foundation for Innovation. We are thankful for the technical advice from research associate Dr.






Kambiz Chizari, technicians Bénédict Besner and Nour Aimene. We would like to specially thank technician Yanik Landry-Ducharme for the help with the hardware of the static puncture test setup.


## 5. References


[1] G. Mayer, Rigid Biological Systems as Models for Synthetic Composites, Science. 310 (2005) 1144–1147. https://doi.org/10.1126/science.1116994.

[2] A.R. Studart, Towards High-Performance Bioinspired Composites, Adv. Mater. 24 (2012) 5024–5044. https://doi.org/10.1002/adma.201201471.

[3] U.G.K. Wegst, H. Bai, E. Saiz, A.P. Tomsia, R.O. Ritchie, Bioinspired structural materials, Nature Mater. 14 (2015) 23–36. https://doi.org/10.1038/nmat4089.

[4] S.E. Naleway, M.M. Porter, J. McKittrick, M.A. Meyers, Structural Design Elements in Biological Materials: Application to Bioinspiration, Adv. Mater. 27 (2015) 5455–5476. https://doi.org/10.1002/adma.201502403.

[5] A.P. Jackson, J.F.V. Vincent, R.M. Turner, The mechanical design of nacre, Proc. R. Soc. Lond. B. 234 (1988) 415–440. https://doi.org/10.1098/rspb.1988.0056.

[6] M. Mirkhalaf, A.K. Dastjerdi, F. Barthelat, Overcoming the brittleness of glass through bio-inspiration and micro-architecture, Nat Commun. 5 (2014) 3166. https://doi.org/10.1038/ncomms4166.

[7] Z. Yin, A. Dastjerdi, F. Barthelat, Tough and deformable glasses with bioinspired cross-ply architectures, Acta Biomaterialia. 75 (2018) 439–450. https://doi.org/10.1016/j.actbio.2018.05.012.

[8] Z. Yin, F. Hannard, F. Barthelat, Impact-resistant nacre-like transparent materials, Science. 364 (2019) 1260–1263. https://doi.org/10.1126/science.aaw8988.







[9]   H. Yazdani Sarvestani, M. Mirkhalaf, A.H. Akbarzadeh, D. Backman, M. Genest, B. Ashrafi, Multilayered architectured ceramic panels with weak interfaces: energy absorption and multi-hit capabilities, Materials & Design. 167 (2019) 107627. https://doi.org/10.1016/j.matdes.2019.107627.

[10]  H. Yazdani Sarvestani, C. Beausoleil, M. Genest, B. Ashrafi, Architectured ceramics with tunable toughness and stiffness, Extreme Mechanics Letters. 39 (2020) 100844. https://doi.org/10.1016/j.eml.2020.100844.

[11]  F. Barthelat, Z. Yin, M.J. Buehler, Structure and mechanics of interfaces in biological materials, Nat Rev Mater. 1 (2016) 16007. https://doi.org/10.1038/natrevmats.2016.7.

[12]  B.L. Smith, J.B. Thompson, N.A. Frederick, J. Kindt, A. Belcher, G.D. Stucky, P.K. Hansma, Molecular mechanistic origin of the toughness of natural adhesives, fibres and composites, Nature. 399 (1999) 761–763.

[13]  G.E. Fantner, E. Oroudjev, G. Schitter, L.S. Golde, P. Thurner, M.M. Finch, P. Turner, T. Gutsmann, D.E. Morse, H. Hansma, P.K. Hansma, Sacrificial Bonds and Hidden Length: Unraveling Molecular Mesostructures in Tough Materials, Biophysical Journal. 90 (2006) 1411–1418. https://doi.org/10.1529/biophysj.105.069344.

[14]  G.E. Fantner, T. Hassenkam, J.H. Kindt, J.C. Weaver, H. Birkedal, L. Pechenik, J.A. Cutroni, G.A.G. Cidade, G.D. Stucky, D.E. Morse, P.K. Hansma, Sacrificial bonds and hidden length dissipate energy as mineralized fibrils separate during bone fracture, Nature Mater. 4 (2005) 612–616. https://doi.org/10.1038/nmat1428.

[15]  A. Nova, S. Keten, N.M. Pugno, A. Redaelli, M.J. Buehler, Molecular and Nanostructural Mechanisms of Deformation, Strength and Toughness of Spider Silk Fibrils, Nano Lett. 10 (2010) 2626–2634. https://doi.org/10.1021/nl101341w.







[16] M.J. Harrington, H.S. Gupta, P. Fratzl, J.H. Waite, Collagen insulated from tensile damage by domains that unfold reversibly: In situ X-ray investigation of mechanical yield and damage repair in the mussel byssus, Journal of Structural Biology. 167 (2009) 47–54. https://doi.org/10.1016/j.jsb.2009.03.001.

[17] S. Cavelier, C.J. Barrett, F. Barthelat, The Mechanical Performance of a Biomimetic Nanointerface Made of Multilayered Polyelectrolytes, Eur. J. Inorg. Chem. 2012 (2012) 5380–5389. https://doi.org/10.1002/ejic.201200626.

[18] Z. Tang, N.A. Kotov, S. Magonov, B. Ozturk, Nanostructured artificial nacre, Nature Mater. 2 (2003) 413–418. https://doi.org/10.1038/nmat906.

[19] E. Ducrot, Y. Chen, M. Bulters, R.P. Sijbesma, C. Creton, Toughening Elastomers with Sacrificial Bonds and Watching Them Break, Science. 344 (2014) 186–189. https://doi.org/10.1126/science.1248494.

[20] J. Wu, L.-H. Cai, D.A. Weitz, Tough Self-Healing Elastomers by Molecular Enforced Integration of Covalent and Reversible Networks, Adv. Mater. 29 (2017) 1702616. https://doi.org/10.1002/adma.201702616.

[21] X. Feng, Z. Ma, J.V. MacArthur, C.J. Giuffre, A.F. Bastawros, W. Hong, A highly stretchable double-network composite, Soft Matter. 12 (2016) 8999–9006. https://doi.org/10.1039/C6SM01781A.

[22] R. Takahashi, T.L. Sun, Y. Saruwatari, T. Kurokawa, D.R. King, J.P. Gong, Creating Stiff, Tough, and Functional Hydrogel Composites with Low-Melting-Point Alloys, Advanced Materials. 30 (2018) 1706885. https://doi.org/10.1002/adma.201706885.






[23] D.R. King, T. Okumura, R. Takahashi, T. Kurokawa, J.P. Gong, Macroscale Double Networks: Design Criteria for Optimizing Strength and Toughness, ACS Appl. Mater. Interfaces. 11 (2019) 35343–35353. https://doi.org/10.1021/acsami.9b12935.

[24] J.P. Gong, Y. Katsuyama, T. Kurokawa, Y. Osada, Double-Network Hydrogels with Extremely High Mechanical Strength, Adv. Mater. 15 (2003) 1155–1158. https://doi.org/10.1002/adma.200304907.

[25] J.-Y. Sun, X. Zhao, W.R.K. Illeperuma, O. Chaudhuri, K.H. Oh, D.J. Mooney, J.J. Vlassak, Z. Suo, Highly stretchable and tough hydrogels, Nature. 489 (2012) 133–136. https://doi.org/10.1038/nature11409.

[26] S. Zou, D. Therriault, F. Gosselin, Spider Web-Inspired Transparent Impact-Absorbing Composite, Social Science Research Network, Rochester, NY, 2020. https://doi.org/10.2139/ssrn.3667137.

[27] S.-Z. Guo, F. Gosselin, N. Guerin, A.-M. Lanouette, M.-C. Heuzey, D. Therriault, Solvent-Cast Three-Dimensional Printing of Multifunctional Microsystems, Small. 9 (2013) 4118–4122. https://doi.org/10.1002/smll.201300975.

[28] R. Passieux, L. Guthrie, S.H. Rad, M. Lévesque, D. Therriault, F.P. Gosselin, Instability-Assisted Direct Writing of Microstructured Fibers Featuring Sacrificial Bonds, Adv. Mater. 27 (2015) 3676–3680. https://doi.org/10.1002/adma.201500603.

[29] S. Chiu-Webster, J.R. Lister, The fall of a viscous thread onto a moving surface: a 'fluid-mechanical sewing machine,' Journal of Fluid Mechanics. 569 (2006) 89. https://doi.org/10.1017/S0022112006002503.






[30] S. Zou, D. Therriault, F.P. Gosselin, Failure mechanisms of coiling fibers with sacrificial bonds made by instability-assisted fused deposition modeling, Soft Matter. 14 (2018) 9777–9785. https://doi.org/10.1039/C8SM01589A.

[31] R.S. Rivlin, A.G. Thomas, Rupture of rubber. I. Characteristic energy for tearing, J. Polym. Sci. 10 (1953) 291–318. https://doi.org/10.1002/pol.1953.120100303.

[32] N.M. Ribe, M. Habibi, D. Bonn, Liquid Rope Coiling, Annu. Rev. Fluid Mech. 44 (2012) 249–266. https://doi.org/10.1146/annurev-fluid-120710-101244.

[33] P.-T. Brun, C. Inamura, D. Lizardo, G. Franchin, M. Stern, P. Houk, N. Oxman, The molten glass sewing machine, Phil. Trans. R. Soc. A. 375 (2017) 20160156. https://doi.org/10.1098/rsta.2016.0156.

[34] C. Creton, M. Ciccotti, Fracture and adhesion of soft materials: a review, Rep. Prog. Phys. 79 (2016) 046601. https://doi.org/10.1088/0034-4885/79/4/046601.

[35] R. Long, C.-Y. Hui, Fracture toughness of hydrogels: measurement and interpretation, Soft Matter. 12 (2016) 8069–8086. https://doi.org/10.1039/C6SM01694D.

[36] Y. Qi, Z. Zou, J. Xiao, R. Long, Mapping the nonlinear crack tip deformation field in soft elastomer with a particle tracking method, Journal of the Mechanics and Physics of Solids. 125 (2019) 326–346. https://doi.org/10.1016/j.jmps.2018.12.018.

[37] S. Lin, C. Cao, Q. Wang, M. Gonzalez, J.E. Dolbow, X. Zhao, Design of stiff, tough and stretchy hydrogel composites via nanoscale hybrid crosslinking and macroscale fiber reinforcement, Soft Matter. 10 (2014) 7519–7527. https://doi.org/10.1039/C4SM01039F.

[38] Z. Wang, C. Xiang, X. Yao, P. Le Floch, J. Mendez, Z. Suo, Stretchable materials of high toughness and low hysteresis, Proc Natl Acad Sci USA. 116 (2019) 5967–5972. https://doi.org/10.1073/pnas.1821420116.







[39] R. Chang, Z. Chen, C. Yu, J. Song, An Experimental Study on Stretchy and Tough PDMS/Fabric Composites, Journal of Applied Mechanics. 86 (2019) 011012. https://doi.org/10.1115/1.4041679.

[40] S. Zou, D. Therriault, F.P. Gosselin, Finite Element Simulation of Microstructured Fibers with Sacrificial Bonds, in: Proceedings of the 2017 International Conference in Aerospace for Young Scientists, Beihang University, Beijing, China, 2017: pp. 34–41.

[41] J. Liu, C. Yang, T. Yin, Z. Wang, S. Qu, Z. Suo, Polyacrylamide hydrogels. II. elastic dissipater, Journal of the Mechanics and Physics of Solids. 133 (2019) 103737. https://doi.org/10.1016/j.jmps.2019.103737.

[42] C. Chen, Z. Wang, Z. Suo, Flaw sensitivity of highly stretchable materials, Extreme Mechanics Letters. 10 (2017) 50–57. https://doi.org/10.1016/j.eml.2016.10.002.

[43] H.R. Brown, A Model of the Fracture of Double Network Gels, Macromolecules. 40 (2007) 3815–3818. https://doi.org/10.1021/ma062642y.

[44] R.O. Ritchie, The conflicts between strength and toughness, Nature Mater. 10 (2011) 817–822. https://doi.org/10.1038/nmat3115.

[45] Q. Wu, S. Zou, F.P. Gosselin, D. Therriault, M.-C. Heuzey, 3D printing of a self-healing nanocomposite for stretchable sensors, J. Mater. Chem. C. 6 (2018) 12180–12186. https://doi.org/10.1039/C8TC02883D.

[46] H. Wei, X. Cauchy, I.O. Navas, Y. Abderrafai, K. Chizari, U. Sundararaj, Y. Liu, J. Leng, D. Therriault, Direct 3D Printing of Hybrid Nanofiber-Based Nanocomposites for Highly Conductive and Shape Memory Applications, ACS Appl. Mater. Interfaces. 11 (2019) 24523–24532. https://doi.org/10.1021/acsami.9b04245.






[47] J. Li, T. Kong, J. Yu, K.H. Lee, Y.H. Tang, K.-W. Kwok, J.T. Kim, H.C. Shum, Electrocoiling-guided printing of multiscale architectures at single-wavelength resolution, Lab Chip. 19 (2019) 1953–1960. https://doi.org/10.1039/C9LC00145J.

[48] J. Tian, J. Li, A. Sauret, T. Kong, X. Wu, Y. Lu, H.C. Shum, Facile Control of Liquid-Rope Coiling With Tunable Electric Field Configuration, Phys. Rev. Applied. 12 (2019) 014034. https://doi.org/10.1103/PhysRevApplied.12.014034.

[49] Y.E. Choe, G.H. Kim, A PCL/cellulose coil-shaped scaffold *via* a modified electrohydrodynamic jetting process, Virtual and Physical Prototyping. (2020) 1–14. https://doi.org/10.1080/17452759.2020.1808269.

[50] I. Liashenko, A. Hrynevich, P.D. Dalton, Designing Outside the Box: Unlocking the Geometric Freedom of Melt Electrowriting using Microscale Layer Shifting, Advanced Materials. 32 (2020) 2001874. https://doi.org/10.1002/adma.202001874.

[51] J.C. Kade, P.D. Dalton, Polymers for Melt Electrowriting, Advanced Healthcare Materials. n/a (n.d.) 2001232. https://doi.org/10.1002/adhm.202001232.





Supplementary Material

# Toughening elastomers via microstructured thermoplastic fibers with sacrificial bonds and hidden lengths


Shibo Zou (邹士博) [a]*, Daniel Therriault [a], Frédérick P. Gosselin [a]

[a] Laboratory for Multiscale Mechanics (LM2), Department of Mechanical Engineering, Research Center for High Performance Polymer and Composite Systems (CREPEC), Polytechnique Montréal, Montréal, QC H3T 1J4, Canada.

* Corresponding author. Email: shibo.zou@polymtl.ca






**Flaw size introduced by the P60 sandpaper indenter**

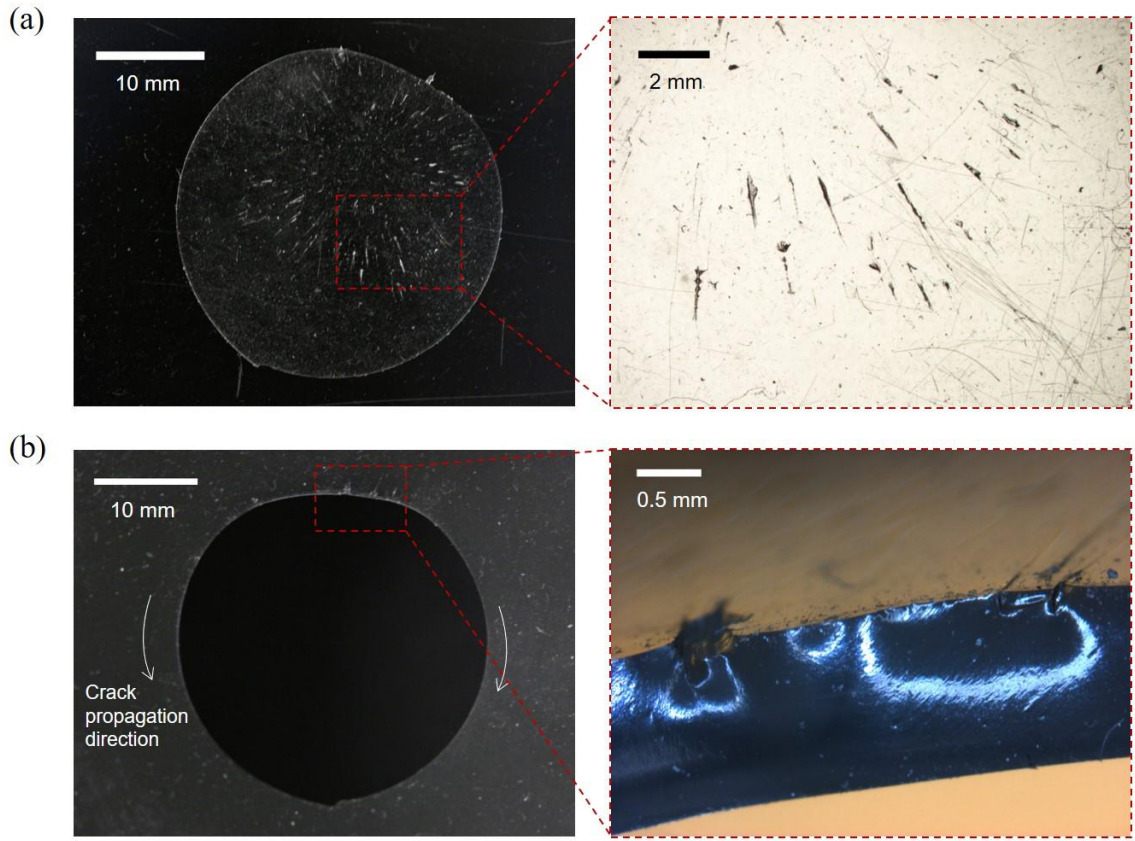

Figure S1. Optical images of the plain PDMS specimen after the static puncture test with the rough indenter covered with P60 sandpaper. (a) The middle piece which fell out of the test specimen after test. An area on the piece is shown at higher magnification in an optical microscopy image. The slender surface cracks were introduced by the P60 sandpaper attached on the indenter. The length and width of the cracks were measured as $1.16 \pm 0.44$ mm and $0.098 \pm 0.046$ mm, respectively. (b) The specimen left with a hole after test. The crack is believed to initiate within the dashed red rectangle, since the opposite side of the circle shows a small crack path mismatch, which could be caused by the encounter of two running cracks. The crack initiation region was characterized in a tilted view under optical microscope. The maximum depth of the surface crack is between 0.3 mm and 0.4 mm.